# Fully Atomic-Layer-Deposited Vertical Complementary FeRAM with Ultra-High $2P_r$ > 100 μC/cm$^2$ and High Endurance > $10^{10}$ cycles

Renhao Xue[a], Ruizhan Yan[a], Mansun Chan[b], Xiwen Liu[a*]

[a] Thrust of Microelectronics, The Hong Kong University of Science and Technology (Guangzhou), Guangzhou, Guangdong, China

[b] Department of Electrical and Computer Engineering, The Hong Kong University of Science and Technology, Kowloon, Hong Kong

*Corresponding author: xiwenliu@hkust-gz.edu.cn

**Abstract**

A limited remanent polarization ($P_r$) in $HfO_2$-based FeRAM remains a key obstacle to density scaling and reliability, while material and process optimizations offer only incremental improvements. This limitation fundamentally originates from the thickness-constrained switchable polarization and the intrinsic polarization ceiling of $HfO_2$-based ferroelectrics. Here, we propose an all-ALD-grown vertical complementary FeRAM (VCF) architecture, in which the top and bottom stacked FeRAM maintain complementary polarization. This complementary dipole configuration converts the readout from a single-layer polarization response into a differential polarization summation, thereby amplifying the effective charge window without increasing the switching field of each individual layer and area overhead. Viewed from top to bottom, an "up–down" polarization pair stores logic '1', whereas a "down–up" pair stores logic '0'. With a complementary polarization write–read scheme, the VCF effectively achieves $2P_r$ > 100 μC/cm$^2$ and maintains $2P_r$ > 90 μC/cm$^2$ after $10^{10}$ switching cycles without electrical breakdown. Robust retention (> $10^4$ s at 85 °C) and strong disturb immunity are demonstrated, with $2P_r$ > 80 μC/cm$^2$ under a V/3 scheme after $10^6$ disturb pulses. Array-level operation is validated in a 5×5 selector-free crosspoint array. The performance enhancement of the VCF arises from the co-optimization of the all-ALD-grown process, device architecture, and operation scheme, enabling high density, a wide memory window, and strong reliability for scalable FeRAM integration.

## Main

Ferroelectric random-access memory (FeRAM) based on $HfO_2$ has emerged as a promising nonvolatile technology for fast and energy-efficient embedded memory applications.[1-3] While the FeRAM crosspoint array has demonstrated excellent high-density potential and back-end-of-line (BEOL) compatibility, it still faces critical challenges in simultaneously achieving high storage density and a large memory window.[4-5] As device dimensions scale down, the stored remanent polarization ($P_r$) per cell decreases rapidly, limiting the sensing margin and array scalability.[6] Recent efforts have focused on materials and process engineering to enhance $2P_r$. Electrode engineering—such as replacing TiN with Nb, Ru, or $RuO_2$—has been shown to increase $2P_r$ in thin HZO films (e.g., $2P_r$ reaching ~ 20 μC/cm² with Ru[7] and ~ 50 μC/cm² with Nb[8]). Epitaxial oxide electrodes (e.g., $SrRuO_3$, $LaNiO_3$) further stabilize the ferroelectric O-phase and yield robust $2P_r$ ~ 45 μC/cm².[9]

Beyond electrode materials optimization, process-level strategies have been employed to enhance $2P_r$ by refining ferroelectric film deposition methods,[10] implementing interface engineering,[11-14] preconditioning electrodes,[14-16] adjusting process compositions,[17-18] and optimizing annealing conditions.[4, 19] Interfacial and seed-layer engineering—such as the insertion of ultrathin $ZrO_2/TiO_2$ interlayers—can effectively enhance ferroelectricity. For example, the insertion of $ZrO_2$ seed layers boosts $2P_r$ from ~ 47 to 56 μC/cm².[11] Rowtu et al. demonstrated that integrating a $TiO_2$ seed with an $Nb_2O_5$ capping layer on La-doped HZO boosts $2P_r$ to ~ 66.5 μC/cm² after $3\times10^6$ cycles.[12-13] Samsung then enhanced $2P_r$ through co-optimization of interface-curing treatments and $P_r$-boosting electrode stacks, raising $2P_r$ to 59 μC/cm².[14] In addition, by incorporating dopants into the HZO layer, $2P_r$ values above 64 μC/cm² are reported by Micron.[16] Intel also reported an anti-ferroelectric FeRAM through compositional engineering and optimized annealing conditions, achieving an effective $2P_r$ exceeding 45 μC/cm$^2$.[18]

In this work, we propose an all-atomic layer deposition (ALD)-grown vertical complementary FeRAM (VCF) architecture. The stacked FeRAM cells maintain complementary polarization, where an "up–down" pair (top–bottom) stores logic '1', and a "down–up" pair stores logic '0'. Analogous to the complementary FET (CFET) concept that has enabled vertical co-integration in advanced logic nodes, the proposed VCF achieves a substantially enlarged memory window while preserving an ultra-compact cell footprint.[20] Using an all-ALD process with ultrathin TiN electrodes and high-quality HZO, the two-layer stacked VCF ensures atomic-scale thickness control, superior conformality, and high-quality interfaces, thereby stabilizing the field distribution and suppress localized degradation during repeated switching. The VCF exhibits $2P_r$ > 100 μC/cm², endurance > $10^{10}$ cycles, robust retention, and reliable operation in a selector-free crosspoint array, offering a promising pathway toward high-density FeRAM integration.

To realize a vertically stacked two-layer FeRAM architecture with high reliability, an all-ALD process is adopted, with plasma $O_2$ and $N_2/H_2$ serving as the oxidant and reactive gases for high-quality TiN and HZO film deposition. Each layer in the stack is precisely thickness-controlled by ALD. Both the TiN electrode and HZO ferroelectric layer are kept ultrathin (~10 nm) to optimize interfacial stress and further suppress roughness, thereby enhancing the reliability of the multi-layer stack. The bottom, middle, and top TiN electrodes are patterned to form the two-layer crosspoint architecture, as shown in Fig. 1(a). A rapid thermal annealing (RTA) step is subsequently performed to crystallize the ferroelectric phase. The stacked architecture, along with cross-sectional HRTEM and EDS mapping of the two-layer FeRAMs, is shown in Fig. 1(b)–1(d), where the vertically separated ferroelectric layers are clearly resolved, confirming the well-defined stacking order.

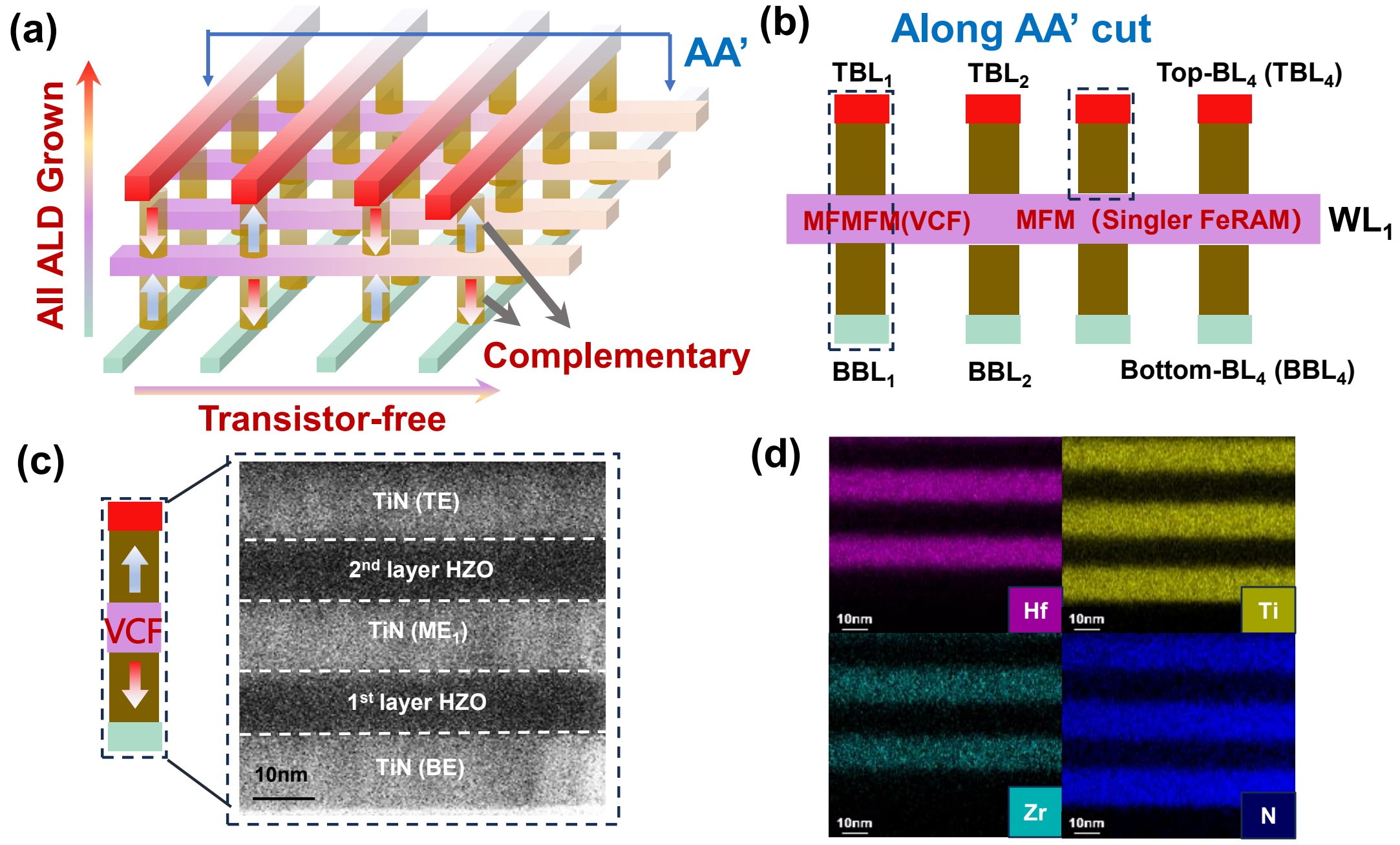

**Fig. 1.** (a) Schematic of an all-ALD-grown vertical complementary FeRAM crosspoint array. Two ferroelectric layers are programmed into complementary polarization. (b) Cross-section along AA' highlighting the alternating MFMFM stacks (M: metal, F: ferroelectric) constituting the array. (c) Cross-sectional TEM image and (d) EDS mapping.

Figure 2 summarizes the operating principle and polarization-voltage (P-V) loop of the proposed VCF. As shown in Fig. 2(a), the VCF adopts a complementary polarization write–read scheme, in which the top and bottom ferroelectric layers are programmed into a complementary polarization. An "up–down" polarization pair (top–bottom) stores logic '1', and a "down–up" pair stores logic '0'. During readout, the charges stored in tier 1 and tier 2 are read out together, yielding a significantly enlarged differential signal window between logic '0' and '1'. The individual P–V loops in Fig. 2(b) show that both ferroelectric layers exhibit well-matched switching behavior, enabling stable and symmetric complementary programming. When operated in the proposed complementary scheme, the response obtained by reading out both layers together yields an effective $2P_r$ window > 100 μC/cm², approximately twice that of a single-layer FeRAM. This enlarged $2P_r$ improves the read margin without any area penalty, highlighting the co-optimization of stack architecture and complementary polarization operation scheme design.

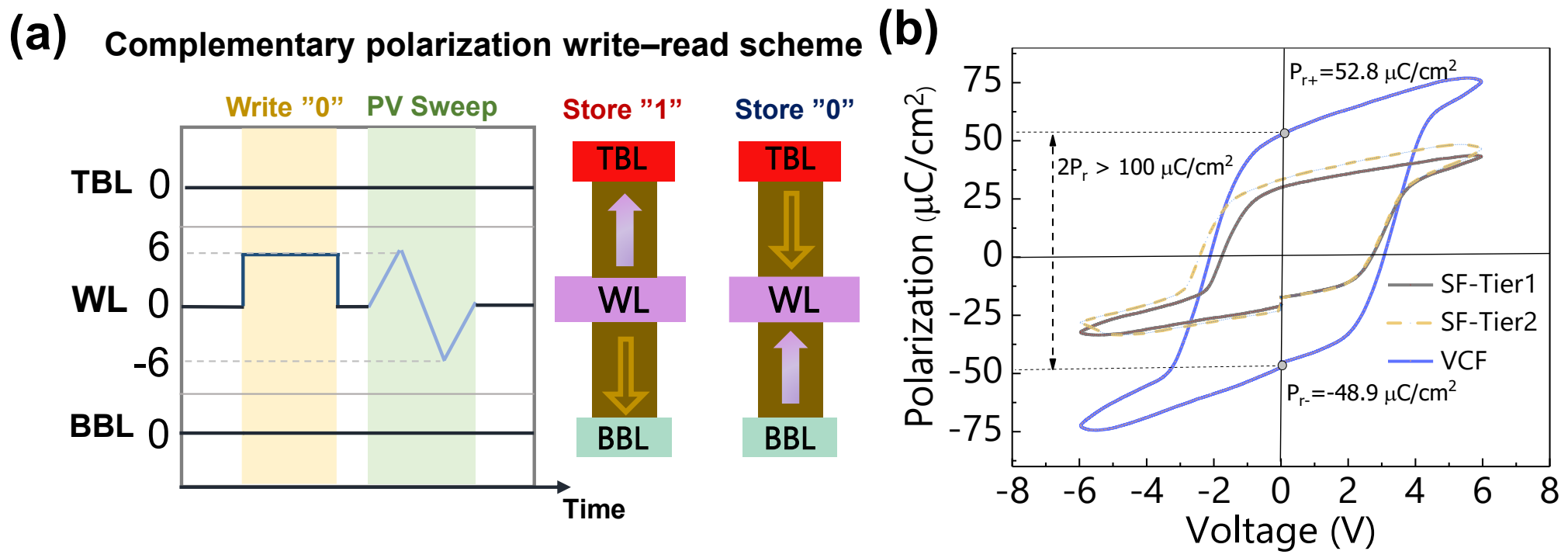

**Fig. 2.** (a) The proposed complementary polarization write–read scheme, illustrating complementary polarization in stacked FeRAM cells, where an "up–down" polarization pair (top–bottom) stores logic '1', and a "down–up" pair stores logic '0'. During readout, both layers are read out together to achieve an ultra-large effective memory window. (b) P–V hysteresis of the individual single-layer FeRAM (SF), showing well-matched switching behavior. With the complementary program/read scheme, the VCF exhibits an enlarged effective window, achieving $2P_r > 100\ \mu C/cm^2$.

Building on the enlarged $2P_r$ demonstrated by the VCF in Fig. 2, Figures 3(a) and 3(b) compare the switching dynamics between a single-layer FeRAM and the VCF. For a target switching polarization (e.g., $2P_r = 40\ \mu C/cm^2$), the VCF reaches the required polarization contour with shorter pulses and lower voltages, indicating faster, lower energy switching and readout. Furthermore, as shown in Fig. 3(c), a target $2P_r$ of ~ 48 $\mu C/cm^2$ that requires ~ 6 V in a single-layer FeRAM can be achieved at only ~ 2.5 V in the VCF, thereby directly reducing read/write power and enabling higher density with improved energy efficiency. Finally, Figure 3(d) shows the measured and fitted I–V characteristics: the VCF exhibits a low disturb parameter $\alpha_{D,\ VCF}$=0.225, comparable to state-of-the-art selector-free FeRAM devices,[21-23] thereby underscoring the competitive disturb immunity of the proposed VCF in crosspoint array operation.

We next evaluate the reliability of the VCF across multiple metrics. As shown in Fig. 4(a), the cell withstands > $10^{10}$ cycles while maintaining $2P_r > 90\ \mu C/cm^2$ without electrical breakdown, demonstrating exceptional cycling robustness and the ability to maintain an ultra-high memory window over prolonged stress—outperforming single-layer FeRAM.[17] Furthermore, the P–V loops in Fig. 4(b) remain highly consistent across different cycle counts, confirming stable write–read operation and negligible fatigue. Disturb testing in Fig. 4(c) further demonstrates the excellent immunity of the VCF array under the V/3 operation scheme:[4] under multiple disturb pulses (~ $10^6$ cycles) stress with an opposite $V_{write/3}$ bias, the effective window exhibits only minor degradation, ensuring its stable operation in selector-free crosspoint arrays. Finally, the retention plot in Fig. 4(d) verifies excellent thermal stability, with the programmed polarization preserved at 85 °C for > $2\times10^4$ s with no observable degradation, confirming the long-term data retention capability of the VCF. Together, these results demonstrate excellent reliability across multiple metrics for large-scale crosspoint deployment.

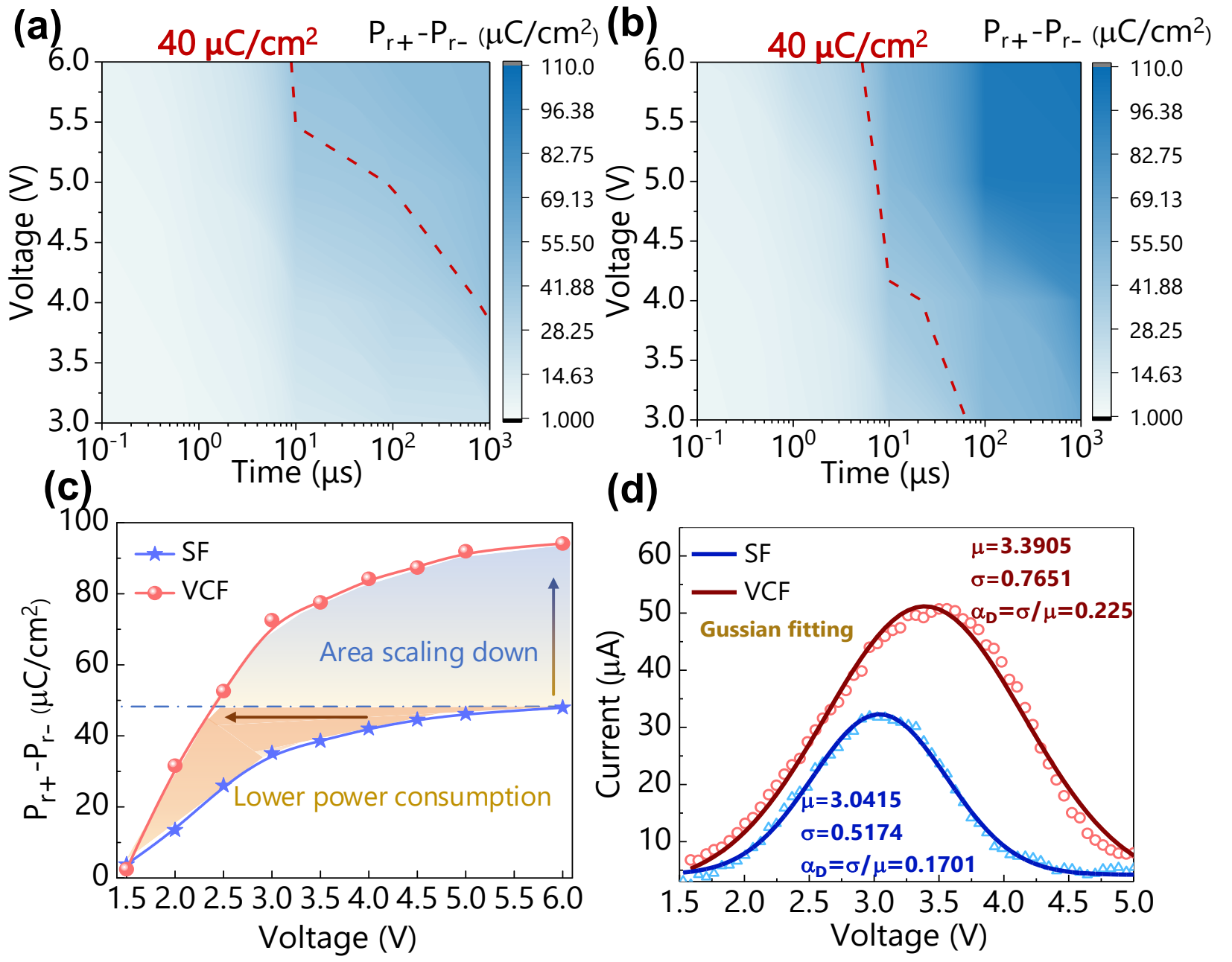

**Fig. 3.** Switching dynamics of the SF (a) and VCF (b). (c) Extracted $2P_r$ versus read voltage. (d) Measured and fitted I–V curves of SF and VCF, yielding a low disturb parameter $\alpha_{D,\,VCF}$=0.225, indicative of good disturb immunity.

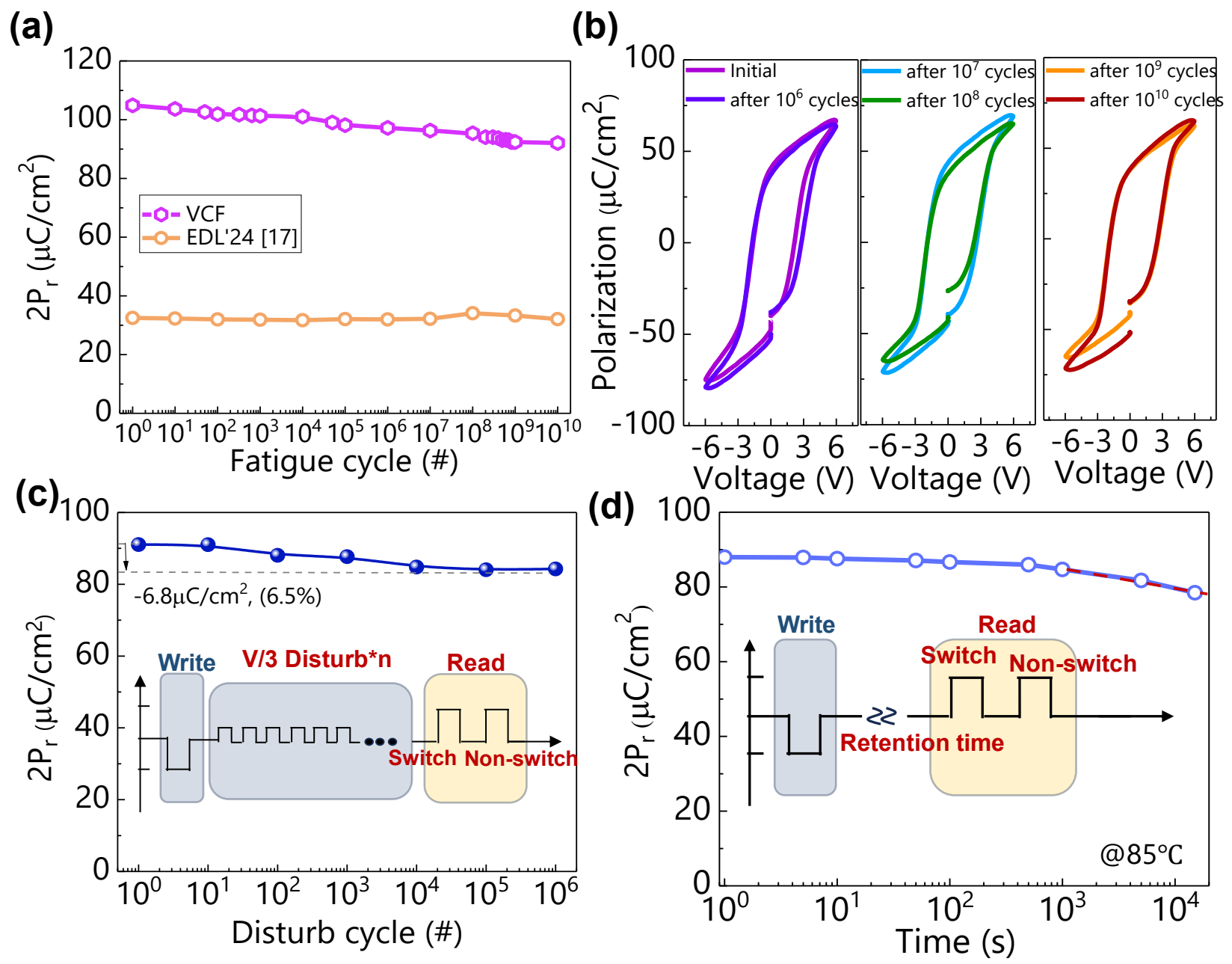

**Fig. 4.** (a) Endurance of the VCF after $10^{10}$ switching cycles. The VCF remains $2P_r$ > 90 μC/cm$^2$ with no electrical breakdown, evidencing robust reliability. (Adapted from Ref. 17. © 2024 IEEE.)

(b) P-V hysteresis loops measured at different cycle counts, showing stable write – read behavior. (c) Measurement of multiple disturb pulses with opposite 1/3 $V_{write}$, showing excellent immunity. (d) Retention characteristics of the VCF at 85.℃

To further confirm the scalability and practicality of the proposed architecture, we fabricated a 5×5 selector-free VCF crosspoint array. Figure 5(a) presents the SEM image of the array. Each crosspoint corresponds to a VCF cell, demonstrating successful selector-free integration. Figure 5(b) shows the output pattern extracted from the switching polarization ($P_{sw}$) of all 25 cells under the complementary operation scheme, where stable and uniform polarization (> 85 μC/cm²) reversal is clearly observed across the array. These results validate the reliable and scalable operation of the VCF concept from single-cell performance to array-level integration.

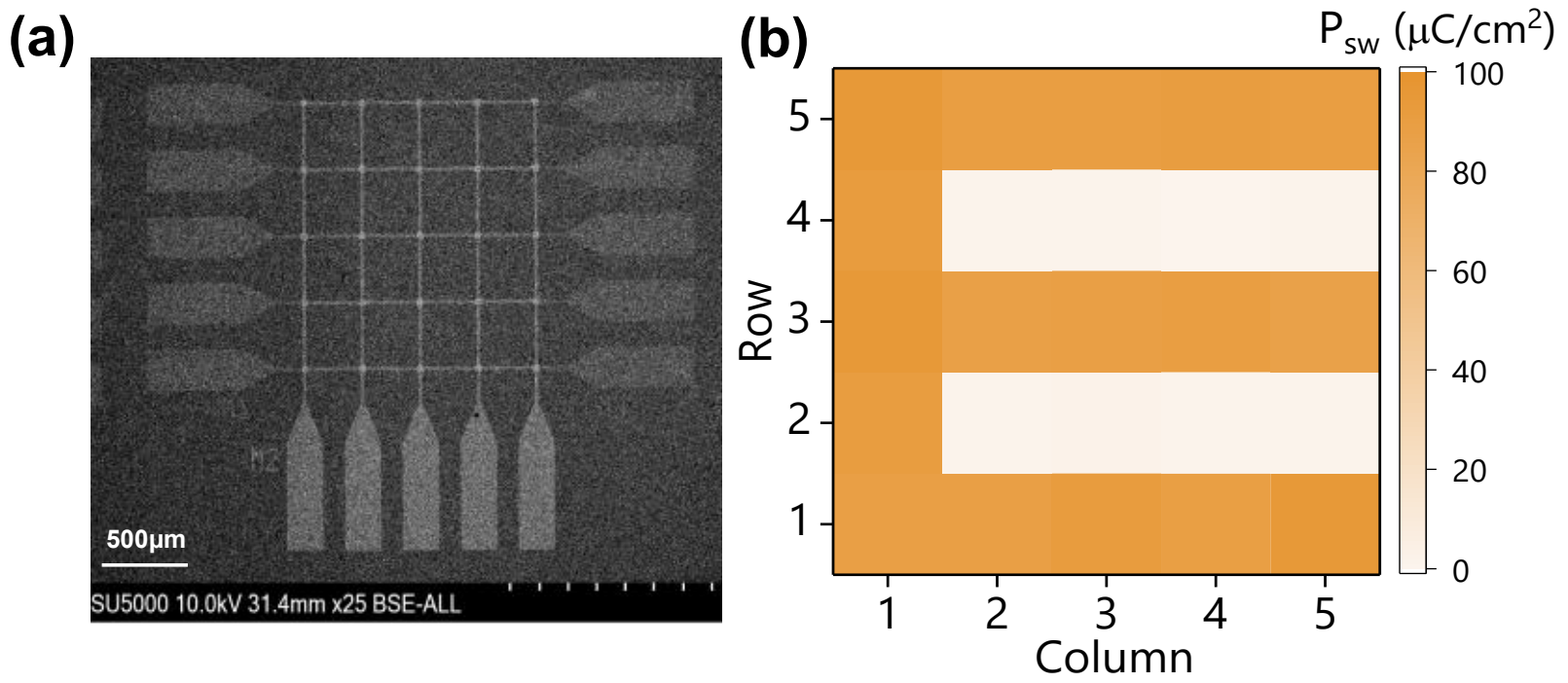

**Fig. 5.** (a) SEM image of the fabricated 5×5 VCF crosspoint array. (b) Output pattern displayed by $P_{sw}$ of the VCF.

Table I further benchmarks the key metrics against state-of-the-art FeRAM, highlighting the superior performance of the proposed VCF.[4, 8, 24-28] This improvement arises from the process, device structure, and operation scheme. All plasma-enhanced ALD growth yields smooth, defect-suppressed interfaces, while the stacked architecture enables high density without additional area overhead. Furthermore, the complementary polarization readout enables an enlarged effective window. This co-optimization from process to architecture enables the VCF to combine high performance with excellent reliability in a compact footprint, establishing strong competitiveness among state-of-the-art FeRAM technologies.

**TABLE I.** Summary of experimental reports of state-of-the-art FeRAM studies

| Reference | Structure | $T_{FE}$ (nm) | $2P_r$ (μC/cm²) | Endurance | Retention |
|---|---|---|---|---|---|
| Fu *et al* [4] | TiN/HZO/W | 10 | 40 | ~$10^{13}$ | 10 years |
| Wu *et al* [8] | Nb/HZO/Nb | 10 | 30 | ~$10^{9}$ | 10 years |
| Zhao *et al* [24] | $n^+$-Si/HZO/TiN | 8 | 84 | ~$10^{9}$ | N.A. |
| Hur *et al* [25] | TiN/HZO/TiN | 10 | 50 | ~$10^{6}$ | 10000 s |
| Shi *et al* [26] | W/HZLO/W | 7 | 43 | ~$10^{9}$ | 10 years |
| Tahara *et al* [27] | TiN/HZO/TiN | 4 | 30 | ~$10^{10}$ | 10 years |
| Chen *et al* [28] | TiN/HZO/TiN | 6 | 18 | ~$10^{12}$ | N.A. |
| **This work** | **TiN/HZO/TiN/HZO/TiN** | **10** | **101** | **>$10^{10}$** | **10 years** |

In summary, leveraging co-design of process improvement, device architecture, and operation scheme, our proposed VCF achieved both ultra-high $2P_r$ > 100 μC/cm² and strong reliability. Array-level measurements further showed stable readout operation. Moreover, the VCF concept can be extended from planar structures to deep-trench architectures, offering an additional orthogonal route to further enhance the effective $2P_r$ and storage density. These results, together with the structural scalability, confirm the strong potential of VCF for reliable high-density FeRAM integration.

**Corresponding Author**

*E-mail: xiwenliu@hkust-gz.edu.cn

**Competing interests**

There are no conflicts to declare.

**Data Availability**

The data that support the conclusions of this study are available from the corresponding authors upon reasonable request.

**ACKNOWLEDGEMENTS**

The authors acknowledge the support from the National Natural Science Foundation of China (No. 62404189), Guangzhou Municipal Science and Technology Project (No. 2025A03J3885, No. 2025A04J4071), Guangdong Scientific Research Platform and Projects for the Higher-educational Institution & Education Science Planning Scheme (2025KTSCX196), and the Start-up fund from the Hong Kong University of Science and Technology (Guangzhou). This work is also supported by ACCESS – AI Chip Center for Emerging Smart Systems under the InnoHK funding of Hong Kong SAR.